\documentclass[pra,10pt,twocolumn,superscriptaddress,showpacs]{revtex4-1}
\usepackage{amsmath}
\usepackage{latexsym}
\usepackage{amssymb}
\usepackage{graphicx}
\usepackage[colorlinks=true, citecolor=blue, urlcolor=blue]{hyperref}
\usepackage{float}
\usepackage{amsfonts}

\newcommand{\ket}[1]{| #1 \rangle}
\newcommand{\bra}[1]{\langle #1 |}
\newcommand{\braket}[2]{\langle #1 | #2 \rangle}
\newcommand{\ketbra}[2]{| #1 \rangle \langle #2 |}
\newcommand{\cosec}{\operatorname{cosec}}

\begin{document}
\title{Measuring higher-dimensional Entanglement}
\author{Chandan Datta}
\email{chandan@iopb.res.in}
\affiliation{Institute of Physics, Sachivalaya Marg,
Bhubaneswar 751005, Odisha, India.}
\affiliation{Homi Bhabha National Institute, Training School Complex, Anushakti Nagar, Mumbai 400085, India.}

\author{Pankaj Agrawal}
\email{agrawal@iopb.res.in}
\affiliation{Institute of Physics, Sachivalaya Marg,
Bhubaneswar 751005, Odisha, India.}
\affiliation{Homi Bhabha National Institute, Training School Complex, Anushakti Nagar, Mumbai 400085, India.}

\author{Sujit K Choudhary}
\email{sujit@iopb.res.in}
\affiliation{Institute of Physics, Sachivalaya Marg,
Bhubaneswar 751005, Odisha, India.}

\begin{abstract}

We study local-realistic inequalities, Bell-type inequalities, for bipartite pure states of finite dimensional 
quantum systems -- qudits. There are a number of proposed Bell-type inequalities for such systems. 
Our interest is in relating the  value of Bell-type inequality function with a
measure of entanglement. Interestingly, we find that one of these inequalities,
the Son-Lee-Kim inequality, can be used to measure entanglement of a pure 
bipartite qudit state and a class of mixed two-qudit states. Unlike the majority 
of earlier schemes in this direction, where number of observables needed to 
characterize the entanglement increases with the dimension of the subsystems,
this method  needs only four observables. We also discuss the experimental
feasibility of this scheme.  It turns out that current experimental set ups can
be used to measure the entanglement using our scheme.
\end{abstract}


\maketitle

\vskip 0.5in

\section{Introduction}
In a much celebrated paper, J. S. Bell established that any realistic interpretation of quantum 
theory is bound to be nonlocal \cite{bell}. Bell established this  by means of an
inequality which is violated by the singlet state of a pair of qubits. Later, it was
shown that some other states also violate this inequality and thus forbid 
local-realistic description for them. All these led to a very natural 
question -- whether this contradiction between quantum theory and local-realism is typical
or it is restricted to some very special cases. Answer to this question came in 1991 when
Nicolas Gisin \cite{gisin} showed that any pure entangled state of a bipartite system
violates a version of Bell's inequality, Clauser-Horne-Shimony-Holt (CHSH) 
inequality  \cite{chsh}.  Thus all pure bipartite states are nonlocal. The maximum quantum violation of this inequality
is $2\sqrt{2}$. It is known as the Tsirelson's bound \cite{tsirelson}, which a two-qubit maximally 
entangled state attains for a particular measurement setting. This leads to another 
interesting question--- is there any relation between violation of Bell's inequality and 
amount of entanglement? In the case of two-qubit entangled states, it can be shown 
that there exist measurement settings for which value of  
the Bell-CHSH operator increases with the entanglement of the state.
Is this also the case for two-qudit ($d > 2$) entangled state? 
We answer this question also in affirmative.

Entanglement in higher dimensional systems is important from both fundamental and practical 
point of view. Higher dimensional entanglement provides important advantages in quantum 
communication than the conventional qubit entanglement. It provides more security against 
eavesdropping in cryptography \cite{durt}, 
it can be used to increase the channel-capacity via superdense-coding \cite{chuang} and is 
more robust against environmental noise \cite{pironio} than the conventional two-qubit 
entanglement.  However, for practical applications of these protocols, experimental 
preparation, detection and quantification of higher dimensional entangled state is of 
crucial importance. The violation of Bell-type inequalities can detect the presence of
entanglement in such systems.
Therefore, Bell-type inequalities in higher dimensional system have generated much 
interest in recent years \cite{peres, cglmp, slk, korean, dada, hplo, jphysA, cdh}.
One of the approaches to
obtain Bell-type inequality in higher dimension employs a projection of multilevel
down to dichotomic one \cite{gisin,peres}.  But sometimes it
is also important to know whether it enables to
probe genuine high-dimensionality or not \cite{cglmp, slk, cerf}. In 2002, Collins, Gisin, 
Linden, Massar, and  Popescu introduced \cite{cglmp} an inequality (henceforth will be 
referred to as CGLMP inequality) which
is known to be the only tight inequality \cite{massanes} for higher dimensional systems. 
But, this inequality is not maximally violated for a maximally entangled state of 
such systems \cite{acin}. Interestingly, in 2006, Son, Lee and Kim  
introduced another set of Bell-type 
inequalities for qudit systems (hereafter, the  SLK inequality) \cite{slk} which is maximally 
violated for a maximally entangled state of two-qudit. We show that for a particular measurement 
setting, the Bell-SLK function (defined below) is zero for product
states. Thus, a nonzero value of this function immediately suggests that the measured 
state is entangled. Interestingly, for this setting, the value of the Bell-SLK function 
increases with the concurrence   \cite{conc1, conc2} of the pure bipartite entangled states. Thus, the SLK 
test can serve as a measure of entanglement for pure bipartite states. The situation about mixed
bipartite states of qudits is complicated. One needs a reliable entanglement measure for a general
mixed state of qudits. Even for general bipartite qubit mixed states, the relation
between entanglement and the value of CHSH function is not known. However, we find
a relation between the entanglement of a class of mixed states, isotropic states, 
and the value of the Bell-SLK function for the state. We further explore the relationship between the entanglement, as measured using negativity, and the value of the Bell-SLK function for a maximally entangled state passed through noisy channels.

The widely adopted method for measuring entanglement of a state is 
the quantum state tomographic reconstruction \cite{tomography}. In this method, a 
complete set of observables is measured on the system to reconstruct its state and thus 
to calculate the entanglement. Though, successfully implemented for lower-dimensional 
systems \cite{lowdimtomo1}, this method is not suitable for systems of higher dimension. 
This is because the number of observables to be measured increases dramatically with the 
dimension of the system \cite{highdimtomo1}. However, there are suggestions to characterize 
the state with less number of observables, but most of these methods are for two-qubit 
systems \cite{lowdimtomo2}. For higher dimensional systems, the alternative suggestions, 
though reduces the number of observables (in comparison to the traditional tomography), 
but the number is still high and increases with the dimension of the 
subsystems \cite{highdimtomo2}. In case of bipartite systems, the number of 
measurements needed is of the order of $d^4$ ($d$ is dimension of each subsystem) \cite{standatom}.
Moreover, the implementation of these observables in 
an experiment is also an issue to be taken proper consideration\cite{obexpfeasible}. 
The measurement of the Bell-SLK function can be a method to measure entanglement of a pure bipartite 
state. Unlike the earlier schemes where number of observables needed 
depends on dimension of subsystems, this scheme needs measurement of only four observables 
to calculate the entanglement of any pure bipartite state. Moreover, this new scheme 
can be implemented in laboratories with the existing technology.

The paper is organized as follows. In Section II, we, discuss the SLK test and  obtain a relation 
between the Bell-SLK function and the concurrence. We extend the study to mixed states in Section III.
The way to measure the Bell-SLK function and hence the concurrence in laboratories is briefly 
discussed in Section IV. We conclude in Section V.  

\section{The SLK inequality and concurrence}

In the SLK test, two far separated observers Alice and Bob, can independently choose one
of the two observables denoted by $A_1$,  $A_2$ for Alice and  $B_1$,
 $B_2$ for Bob. Measurement outcomes of the  observables are elements of the set, 
 $V=\{1,\omega,\cdots,\omega^{d-1}\}$, where $\omega=\exp{(2 \pi i/d)}$. 
In a variant of SLK  inequality \cite{korean}, the Bell-SLK function, $I_{SLK}$, is given by
\begin{eqnarray}
 I_{SLK}&=&\frac{1}{\sqrt{2}}\sum\limits_{n=1}^{d-1}\big(\omega^{-n/4}C_{1,1}^n+\omega^{-3n/4}C_{2,1}^n\nonumber\\&&
+\omega^{n/4}C_{1,2}^n+\omega^{-n/4}C_{2,2}^n\big)+c.c.,
\label{Son_Correl}
\end{eqnarray}
$\omega=exp(2\pi i/d)$, c.c. is for complex conjugate, 
$C_{a,b}^{n}=\langle A_a^{n} B_b^{n}\rangle$.
 The assumption of local-realism implies 
 $I_{SLK} \leq I_{SLK}^{\mbox{max}}\mbox{(LR)}$, 
where $I_{SLK}^{\mbox{max}}\mbox{(LR)}=\frac{1}{\sqrt{2}}\Big(3\cot\frac{\pi}{4d}-\cot\frac{3\pi}{4d} \Big)-2\sqrt{2}$
\cite{korean}.
By using Fourier transformation, we write Bell-SLK function in joint probability space as \cite{lcl,korean}
\begin{eqnarray}
 I_{SLK}&=&\sum\limits_{\alpha=0}^{d-1}f(\alpha)[P(A_1=B_1+\alpha)\nonumber\\&&
+P(B_1=A_2+\alpha+1)+P(A_2=B_2+\alpha)\nonumber\\&&+P(B_2=A_1+\alpha)],
\label{Son_In_probability}
\end{eqnarray}
where sums inside the probabilities are modulo $d$ sums, and
\begin{equation}
 f(\alpha)=\frac{1}{\sqrt{2}} \Big(\cot[\frac{\pi}{d}(\alpha+\frac{1}{4})]-1 \Big).
\label{form_falpha}
\end{equation}

We now calculate the value of the Bell-SLK function
for an arbitrary pure two-qudit state $\ket{\psi}=\sum_ic_i\ket{ii}$ and for the 
measurement settings originally given in \cite {dagomir}. 
The nondegenerate eigenvectors of the operators 
$\hat{A_a}$, $a=1,2,$ and $\hat{B_b}$, $b=1,2,$ are respectively
\begin{eqnarray}\label{settings}
 \ket{k}_{A,a}=\frac{1}{\sqrt{d}}\sum\limits_{j=0}^{d-1}\omega^{(k+\delta_a)j}\ket{j},\nonumber\\
\ket{l}_{B,b}=\frac{1}{\sqrt{d}}\sum\limits_{j=0}^{d-1}\omega^{(-l+\epsilon_b)j}\ket{j},
\label{eigenvector}
\end{eqnarray}
where $\delta_1=0, \delta_2=1/2, \epsilon_1=1/4$ and $\epsilon_2=-1/4.$ 
The joint probabilities in (\ref{Son_In_probability}) can be calculated as
 \begin{eqnarray}
&&P(A_1=B_1+\alpha)=\frac{1}{d}\sum\limits_{p,q=0}^{d-1}c_pc_q\omega^{(\alpha+1/4)(p-q)},\nonumber\\&&
P(B_1=A_2+\alpha+1)=\frac{1}{d}\sum\limits_{p,q=0}^{d-1}c_pc_q\omega^{-(\alpha+1/4)(p-q)},\nonumber\\&&
P(A_2=B_2+\alpha)=\frac{1}{d}\sum\limits_{p,q=0}^{d-1}c_pc_q\omega^{(\alpha+1/4)(p-q)}, \nonumber\\&&
P(B_2=A_1+\alpha)=\frac{1}{d}\sum\limits_{p,q=0}^{d-1}c_pc_q\omega^{-(\alpha+1/4)(p-q)}.
\end{eqnarray}
Putting these probabilities in (\ref{Son_In_probability}), we get  
\begin{eqnarray}
 I_{SLK}=\frac{4}{d}\sum\limits_{\alpha=0}^{d-1}f(\alpha)\sum\limits_{p,q=0}^{d-1}c_pc_q\omega^{(\alpha+1/4)(p-q)}.
\label{expectation_value}
\end{eqnarray}
From the identity $\sum\limits_{k=0}^{d-1}(-1)^k\cot
(\frac{2k+1}{4d})\pi=d$ \cite{Hassan}, we can obtain another identity $\sum\limits_{k=0}^{d-1}\cot(\frac{4k+1}{4d})\pi=d$.
Therefore, we get $\sum\limits_
{\alpha=0}^{d-1}f(\alpha)=0$.  We can then rewrite  (\ref{expectation_value}) as

\begin{eqnarray}
 I_{SLK}&&=\frac{4}{d}\sum\limits_{\alpha=0}^{d-1}f(\alpha)\sum_{\substack{p\neq q\\ p>q}}2c_pc_q\cos\Big(\frac{2\pi}{d}
(\alpha+\frac{1}{4})(p-q)\Big)\nonumber\\&&
=\frac{4}{d}\sum\limits_{\alpha=0}^{d-1}\frac{1}{\sqrt{2}} \Big(\cot[\frac{\pi}{d}(\alpha+\frac{1}{4})]-1 \Big)\nonumber\\&&
\quad \quad \sum_{\substack{p\neq q\\ p>q}}2c_pc_q\cos\Big(\frac{2\pi}{d}
(\alpha+\frac{1}{4})(p-q)\Big).
\label{expectation_value_1}
\end{eqnarray}
To evaluate it further, we now need to find following two sums
\begin{equation}\label{sum1}
 \sum\limits_{\alpha=0}^{d-1}\cos\Big(\frac{2\pi m }{d}(\alpha+\frac{1}{4})\Big),
\end{equation}
and
\begin{equation}\label{sum2}
 \sum\limits_{\alpha=0}^{d-1}\cos\Big(\frac{2\pi m }{d}
(\alpha+\frac{1}{4})\Big)\cot\Big(\frac{\pi}{d}(\alpha+\frac{1}{4})\Big),
\end{equation}
where we have replaced $p-q$ by $m$ (an integer).

Using trigonometrical identity in \cite{Gradshteyn}, we obtain
\begin{eqnarray}
&&\sum\limits_{\alpha=0}^{d-1}\cos\Big(\frac{2\pi m}{d}(\alpha+\frac{1}{4})\Big)\nonumber\\&&
=\cos\Big(\frac{\pi m}{2d}+
\frac{(d-1)\pi m}{d}\Big)\sin\pi m \cosec\frac{\pi m}{d}\nonumber\\&&=0, 
\label{identity_1}
\end{eqnarray}
i.e, the first sum (\ref{sum1}) is equal to zero.

To calculate the sum (\ref{sum2}), we will need two sums that are not available
in well known mathematics handbooks. Motivated by this, we have recently obtained
a number of trigonometric functions sums  \cite{cpsums}. Here we need corollaries of the
theorems in \cite{cpsums}. So we only indicate how to obtain one sum. Other is obtained
in the similar manner.

{\em \bf Proposition:} With $a$ and $k$ being positive integer such that $a < k$, and $ 0 < b < 1$,
\begin{eqnarray}\label{identity1}
 &&\sum\limits_{j=0}^{k-1} \cos{\frac{2\pi aj}{k}}\cot{(\frac{\pi j}{k}+\pi b)} \nonumber\\
 &&=k\cos{[b(2a-k)\pi]}\cosec{(bk\pi)},
\end{eqnarray}

{\em \bf Proof:} To obtain the sum, we will use the method of residues. We take the 
complex function as \cite{Berndt} 
\begin{equation}
	g_{1}(z)=\frac{e^{2\pi iaz}\cot{(\pi z+\pi b)}}{e^{2\pi ikz}-1}-\frac{e^{-2\pi iaz}\cot{(\pi z+\pi b)}}{e^{-2\pi ikz}-1}.
\end{equation}
Let us now consider the integral $\frac{1}{2\pi i}\int_C g_{1}(z)dz$, where $C$ is a contour which is
positively oriented indented rectangle with vertices at $\pm iR$ and $1\pm iR$, with 
$R>\epsilon$, and with semicircular indentations of radius $\epsilon < b$ to the 
left of both $0$ and $1$. Since the period of $g_{1}(z)$ is $1$, 
the integrals along the intended vertical sides of $C$ cancel. 
Here we have taken $a$ such that $0<a<k$.
$g_{1}(z)$ tends to $0$ uniformly for $0 \leqslant x \leqslant 1$ as $\lvert y\rvert\ \rightarrow \infty$.
Hence, $\frac{1}{2\pi i}\int_C g_{1}(z)dz=0$. We, next, calculate the residue at different poles of $g_{1}(z)$. 
$g_{1}$ has simple poles at $z=0$ and $z=j/k, 1\leqslant j \leqslant k-1$. The corresponding residues can be calculated as 
\begin{eqnarray}
	&&\mbox{Res}(g_{1},0)=\frac{1}{\pi i k} \cot(\pi b),\nonumber\\
	&&\mbox{Res}(g_{1},j/k)=\sum\limits_{j=1}^{k-1} \frac{1}{\pi i k}\cos{\frac{2\pi aj}{k}}\cot{(\frac{\pi j}{k}+\pi b)}.
\end{eqnarray}
There is another pole inside the contour -- simple pole at $z=-b+1$. The residue at this point is
\begin{equation}
	\mbox{Res}(g_{1},-b+1)=\frac{i}{\pi}\cos{[b(2a-k)\pi]}\cosec{(bk\pi)}.
\end{equation}
 As the integral is zero, the sum of the residues must be zero. We, thus, get
\begin{eqnarray}
 &&\sum\limits_{j=0}^{k-1} \cos{\frac{2\pi aj}{k}}\cot{(\frac{\pi j}{k}+\pi b)} \nonumber\\
 &&=k\cos{[b(2a-k)\pi]}\cosec{(bk\pi)}.
\end{eqnarray}

This completes the proof \cite{footnote}.

 Similarly, we can obtain, with $a$ and $k$ being positive integer such that $a < k$, and $ 0 < b < 1 $.
\begin{eqnarray}\label{identity2}
 &&\sum\limits_{j=0}^{k-1} \sin{\frac{2\pi aj}{k}}\cot{(\frac{\pi j}{k}+\pi b)} \nonumber\\
 &&=-k\sin{[b(2a-k)\pi]}\cosec{(bk\pi)}.
\end{eqnarray}

Using above two results and sum and difference formula for cosines, we find.
\begin{equation}
 \sum\limits_{\alpha=0}^{d-1}\cos\Big(\frac{2\pi m}{d}(\alpha+\frac{1}{4})\Big)\cot\Big(\frac{\pi}{d}
(\alpha+\frac{1}{4})\Big)=d.
\label{identity_3}
\end{equation}

  This sum is remarkably simple. We note that the value of this sum is independent of $m$. This is
  crucial in relating the value of the Bell-SLK function and entanglement.
    
Eqs. (\ref{form_falpha}), (\ref{expectation_value_1}), (\ref{identity_1}) and (\ref{identity_3}) together imply 
\begin{equation}
 I_{SLK}=4\sqrt{2}\sum_{\substack{p\neq q\\ p>q}}c_pc_q.
\label{expectation_value_2}
\end{equation}

This sum is proportional to the concurrence of the state. The concurrence, $\mathbb{C}$, for a 
two-qudit pure state is defined as  \cite{conc1} 
\begin{equation}\label{concurrence}
\mathbb{C}=\sum_{\substack{p\neq q\\ p>q}}c_pc_q
\frac{2}{d-1}. 
\end{equation}
This is a generalization of the concurrence for a system of two qubits \cite{conc2}.
Using this, we finally get 
\begin{equation}
 I_{SLK}=2\sqrt{2}(d-1)\mathbb{C}.
\label{concurrence_entanglement}
\end{equation}
Thus, we obtain an interesting relation between concurrence and the value of the Bell-SLK function
for a particular measurement setting. Value of this function is zero for product states 
 whereas it increases linearly with the concurrence for
pure entangled states \cite{bellfootnote}. This gives a way to measure the entanglement of a pure entangled state.
The entanglement can be calculated by measuring the Bell-SLK function for the state for the above, given in  (\ref{eigenvector}), measurement setting.

\section{Case of Mixed States}

\subsection{Isotropic States}

As discussed earlier, the case of mixed states even for bipartite qubit systems is far from simple.
In the case of mixed bipartite qudit states, one needs a proper measure of entanglement.
For bipartite mixed states, the concurrence $\mathbb{C}(\rho)$ is calculated by the `convex-roof extension' of the pure-state
concurrence, i.e., by minimizing the average value of the concurrence (mentioned in Eq. (\ref{concurrence}))
over all ensemble decompositions of the mixed state $\rho$:
\begin{equation}\nonumber
\mathbb{C}(\rho)=\inf\limits_{\{p_i, \ket{\psi_i}\}} \Bigg\{\sum_{i}p_i \mathbb{C}(\ket{\psi_i})\Bigg|\sum_i p_i \ket{\psi_{i}}\bra{\psi_{i}}=\rho \Bigg\}.
\end{equation}
Interestingly, in \cite{rungta_caves}, this concurrence has been shown to be an entanglement monotone. However,
due to extremization involved in the calculation, the concurrence in the closed form has been obtained only for a special class of mixed state, namely, for the isotropic states. For a two-qudit system, isotropic states are convex mixtures of the maximally entangled state,
\begin{equation}\label{maximally entangled}
\ket{\Psi^+}=\frac{1}{\sqrt{d}}\sum\limits_{j=0}^{d-1}\ket{jj},
\end{equation}
with a maximally mixed state $I=I \otimes I/d^2$. These states can be written as
\begin{equation}\label{Isotropic state}
\rho_F=\frac{1-F}{d^2-1}\big(I-\ket{\Psi^+}\bra{\Psi^+}\big)+F\ket{\Psi^+}\bra{\Psi^+},
\end{equation} 
where $F$ is the fidelity of $\rho_F$ and $\ket{\Psi^+}$ satisfying  $0\leqslant F \leqslant 1$. These states are separable for $F\leqslant 1/d$ \cite{rungta_horodecki}. 

In the following, we calculate the Bell-SLK function for these states. The Bell-SLK function $I_{SLK}$ (given in Eq. (\ref{Son_In_probability})) consists of four probabilities. Here, we calculate one of the probabilities, $P(A_1=B_1+\alpha)$, other probabilities can be calculated similarly.
\begin{eqnarray}\label{Tr prob}
&&P(A_1=B_1+\alpha)\nonumber\\&&=\mbox{Tr}\big[\hat{P}(A_1=B_1+\alpha)\rho_F\big]  \nonumber\\
&&=\frac{1-F}{d^2-1}\mbox{Tr}\big[\hat{P}(A_1=B_1+\alpha)I\big]+\nonumber\\
&&\frac{d^2F-1}{d^2-1}\mbox{Tr}\big[\hat{P}(A_1=B_1+\alpha)\ket{\Psi^+}\bra{\Psi^+}\big];
\end{eqnarray}
where $\hat{P}(A_1=B_1+\alpha)$ stands for appropriate projector.
The first part of the above sum can be calculated as
\begin{eqnarray}\label{Tr identity}
&&\mbox{Tr}\big[\hat{P}(A_1=B_1+\alpha)I\big]\nonumber\\
&&=\frac{1}{d^2}\sum_{i,j,p,q,l,m,n}
\omega^{(l+\alpha)(i-j)}
\omega^{(-l+1/4)(p-q)}\nonumber\\&&
\braket{m}{i}\braket{j}{m}\braket{n}{p}\braket{q}{n}\nonumber\\
&&=d
\end{eqnarray}
whereas the second part as
\begin{eqnarray}\label{Tr maximally entangled}
&&\mbox{Tr}\big[\hat{P}(A_1=B_1+\alpha)\ket{\Psi^+}\bra{\Psi^+}\big]\nonumber\\
&&=\bra{\Psi^+}\hat{P}(A_1=B_1+\alpha)\ket{\Psi^+}\nonumber\\
&&=\frac{1}{d^3}\sum_{i,j,l,r,s,p,q}\omega^{(l+\alpha)(r-s)}\omega^{(-l+1/4)(p-q)}\nonumber\\
&&\braket{i}{r}\braket{i}{p}\braket{s}{j}\braket{q}{j}\nonumber\\
&&=\frac{1}{d^2}\sum_{i,j}\omega^{(\alpha+1/4)(i-j)}.
\end{eqnarray}
The other three joint probabilities when calculated, come out to be equal to the probability calculated above. The Bell-SLK function, $I_{SLK}$ can now be obtained, by putting for these probabilities in Eq. (\ref{Son_In_probability}), as:
\begin{eqnarray}\label{bellfunctionisotropic0}
 I_{SLK}&=&\bigg(\frac{1-F}{d^2-1}\bigg)4d\sum\limits_{\alpha=0}^{d-1}f(\alpha)+\nonumber\\
&&\bigg(\frac{d^2F-1}{d^2-1}\bigg)\frac{4}{d^2}\sum\limits_{i,j,\alpha=0}^{d-1}f(\alpha)\omega^{(\alpha+1/4)(i-j)}.
\end{eqnarray}
  
Using the fact that $\sum\limits_{\alpha=0}^{d-1}f(\alpha)=0$ (This has been shown in Section II.), the Bell-SLK function reads  

\begin{equation}\label{bell function isotropic}
I_{SLK}=\bigg(\frac{d^2F-1}{d^2-1}\bigg)\frac{4}{d^2}\sum\limits_{i,j,\alpha=0}^{d-1}f(\alpha)\omega^{(\alpha+1/4)(i-j)}.
\end{equation}
Proceeding now in a manner similar as in section (II), we get the Bell-SLK function as 
\begin{equation}\label{bell function isotropic state}
I_{SLK}=\frac{2\sqrt{2}}{d+1}(d^2F-1).
\end{equation}
The concurrence $\mathbb{C}(\rho_F)$ for isotropic states (with some normalization) has been calculated in \cite{rungta_caves} as
\begin{equation}\label{isotropic concurrence}
 \mathbb{C}(\rho_F)  =
    \begin{cases}
      0, & F\leqslant 1/d, \\
     \frac{dF-1}{d-1} , & 1/d\leqslant F \leqslant 1.
    \end{cases}
\end{equation}
Using Eq.(\ref{isotropic concurrence}) in Eq. (\ref{bell function isotropic state}), we  can be rewrite the later
equation to read as
\begin{equation}\label{bell function isotropic concurrence}
I_{SLK}=
\begin{cases}
 \frac{2\sqrt{2}}{d+1}(d^2F-1), & F\leqslant 1/d, \\
\frac{2\sqrt{2}}{d+1}\big((d-1)(d\mathbb{C}(\rho_F)+1)\big), & 1/d\leqslant F \leqslant 1.
\end{cases}
\end{equation}
Thus, we get an interesting relation between the Bell-SLK function and the concurrence for isotropic states.
It can easily be checked that for $F\leqslant 1/d$, i.e. for the separable isotropic states, the value of the Bell-SLK function is upper bounded by the $\frac{2\sqrt{2}}{d+1}(d-1)$. A value larger than this bound, immediately suggests that the isotropic states undergoing the said Bell-SLK measurements are entangled and their entanglement increases with the value of the Bell-SLK function.   

\subsection{Maximally Entangled state through a noisy channel}

In this section, we will consider another class of mixed states that are
obtained when particles in a maximally entangled state pass through noisy 
channels. This is often a real laboratory situation.
As we have already mentioned there is no closed from of concurrence for a 
general mixed state of two qudits; so here we use another measure of entanglement, 
called negativity \cite{vidal}. The negativity of a state $\rho$ is defined as 
\begin{equation}\label{negativity}
\mathcal{N}(\rho)=\frac{||\rho^{T_B}||_1-1}{d-1},
\end{equation}  
where $||\cdot||_1$ represents the trace norm and $\rho^{T_B}$ is the partial transpose of 
the state $\rho$ with respect to the subsystem $B$. For a pure state $\ket{\psi}=\sum_ic_i\ket{ii}$ 
the negativity can be written as \cite{soojoon}
\begin{equation}\label{negativity pure}
\mathcal{N}\big(\ketbra{\psi}{\psi}\big)=\sum_{\substack{p\neq q\\ p>q}}c_pc_q
\frac{2}{d-1},
\end{equation}
where $c_i$ are the Schmidt coefficients. So it is same as concurrence for a pure state. Therefore, we can 
write (\ref{concurrence_entanglement}) also as
\begin{equation}\label{slk negativity}
I_{SLK}=2\sqrt{2}(d-1)\mathcal{N}.
\end{equation}
 Let's say a party prepares 
a maximally entangled state and sends it to two distant parties by some noisy channels. 
The state will no longer be a pure state. Since we can compute negativity for a mixed state,
we can explore the relationship between negativity and the value of the Bell-SLK function. 
We can also find how robust is the measurement of entanglement of a pure state using the
Bell-SLK function. We will study this situation in $d=3$ taking two well known 
channels -- the amplitude damping and the phase damping channels.  

\subsubsection{Amplitude damping} 

Let us consider a maximally entangled state $\ket{\psi}=\frac{1}{\sqrt{3}}(\ket{00}+\ket{11}+\ket{22})$. 
Two qutrits are sent to distant parties through amplitude damping channels. 
For a qutrit, amplitude damping 
channel can be represented in terms of Kraus operators as \cite{ramzan}

\quad\quad\quad\quad\quad$K_0=\begin{pmatrix}
1 & 0 & 0 \\
0 & \sqrt{1-p} & 0 \\
0 & 0 & \sqrt{1-p} 
\end{pmatrix}$,\\

\begin{figure}[H]
\centering
\includegraphics[scale=0.65]{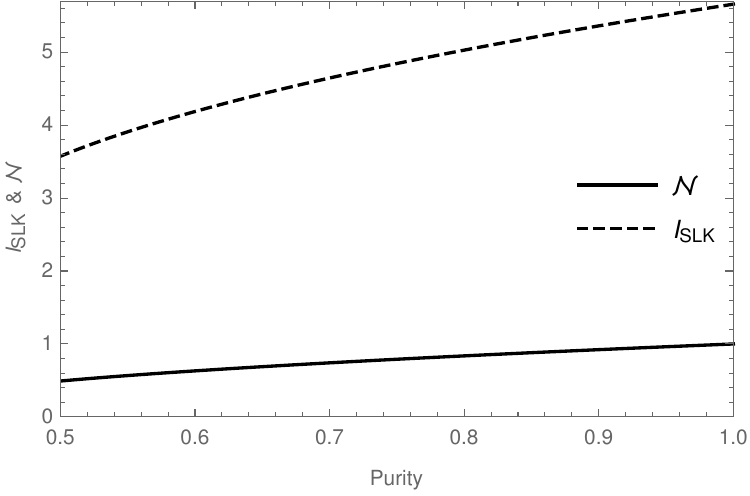}
\caption{Change in the Bell-SLK function and negativity with purity for a maximally entangled two-qutrit state passed through the amplitude damping channel.}
\label{amplitude}
\end{figure}

\quad\quad\quad$K_1=\begin{pmatrix}
0 & \sqrt{p} & 0 \\
0 & 0 & 0 \\
0 & 0 & 0 
\end{pmatrix}$,
$K_2=\begin{pmatrix}
0 & 0 & \sqrt{p} \\
0 & 0 & 0 \\
0 & 0 & 0 
\end{pmatrix}$,  \\
where $p$ is the channel parameter. For simplicity we take same channel on both sides. 
We find that, without noise,
the value of $I_{SLK}$ is $5.657$ and negativity is $\mathcal{N}=1$. At
$90\%$ purity (purity varies with the channel parameter, $p$) $I_{SLK}=5.359$ and negativity is $\mathcal{N}=0.922$. So 
the value of negativity is about $8\%$ lower. However, it turns out that
there is still a relationship between the value of the Bell-SLK function 
and entanglement for such states.
From the value of the Bell-SLK function we can infer the entanglement 
of the state in terms of negativity. This is clear from \figurename{\ref{amplitude}} 
that looking at the $I_{SLK}$ curve, we can find state's negativity at any purity. 
 
\subsubsection{Phase damping}

Let us now consider the case when the two qutrits in a maximally
entangled state pass through phase damping channels separately.
The Kraus operators for the phase damping channel are \cite{ramzan}

\begin{figure}[H]
\centering
\includegraphics[scale=0.65]{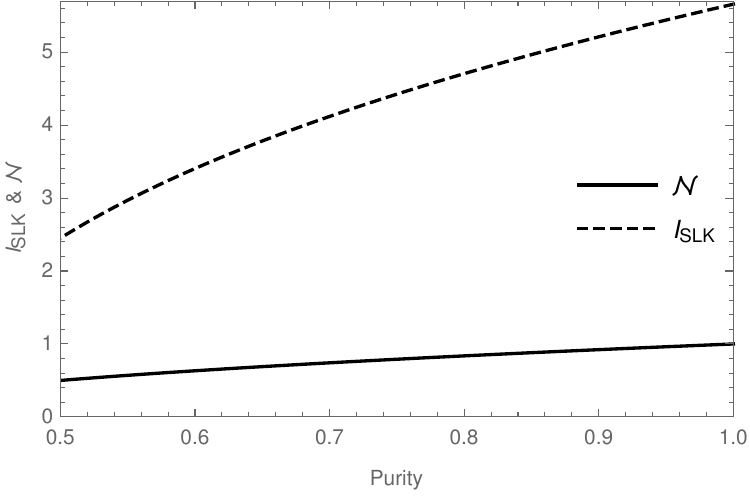}
\caption{Change in the Bell-SLK function and negativity with purity for a maximally entangled two-qutrit state passed through the phase damping channel.}
\label{phase}
\end{figure}

\quad$K_0=\sqrt{1-p}\begin{pmatrix}
1 & 0 & 0 \\
0 & 1 & 0 \\
0 & 0 & 1 
\end{pmatrix}$ and 
$K_1=\sqrt{p}\begin{pmatrix}
1 & 0 & 0 \\
0 & \omega & 0 \\
0 & 0 & \omega^2 
\end{pmatrix}$, \\
where $\omega=e^{\frac{2 \pi i}{3}}$ and $p$ is the channel parameter. We do the same analysis 
as for the amplitude damping channel. 
At $90\%$ purity $I_{SLK}=5.209$ and $\mathcal{N}=0.922$. Both values decrease by about
$8\%$, as compared to the starting pure state. However, as before, 
from \figurename{\ref{phase}}, we notice an interesting relationship between entanglement
and the value of the Bell-SLK function. If we measure the $I_{SLK}$ value for the state,
we can easily determine the entanglement of the state in terms of negativity. 

\section{An Experimental scheme}

 Interestingly, this SLK test can be performed in laboratories with the present day's technology \cite{dada,hplo}. One technique to
 encode a state of a qudit is to use
the orbital angular momentum (OAM) states of photons \cite{zeilinger1}. Higher dimensional 
bipartite entanglement is generated through spontaneous parametric down conversion (SPDC) 
\cite{zeilinger2, dada}. In \cite{dada}, Dada {\em et al.} have employed the same measurement setting as in Eq. (\ref {settings}) to obtain the violation of CGLMP inequality for bipartite qudit systems with dimensions up to twelve.  
One can use the same experimental set up to measure the Bell-SLK function
 (instead of the CGLMP function as done in \cite{dada}) in order to find the amount of entanglement present 
 in a pure bipartite state. 
In this case, four observables are to be measured. However, in this experimental set up, the measurement of each
observable requires $(d-1)$ experimental settings. The number of settings varies linearly
with respect to the  dimension $d$. In principle, it might be possible to reduce the number
of settings to measure an observable with $d$ outcomes. In \cite{hplo}, Lo {\em et al.} employ
a different experimental set up.They simulate qudits using multiple pairs of polarization-entangled photons.  
They also measure four observables with $(d-1)$ experimental settings for each observable and
demonstrate the violation of CGLMP inequality up to $d = 16$.
Though violation of the CGLMP inequality can detect the presence of entanglement, but this violation cannot be used to measure the amount of entanglement present in the bipartite state, at least, for the employed setting. 
This is because, for these settings, CGLMP inequality  is not maximally violated for a maximally entangled state of two qudits
\cite{acin}. It is also known \cite{acin} that CGLMP inequality is violated maximally by a partially entangled state.
Therefore, it is unlikely that CGLMP function measurement can help in measuring the amount
of entanglement in a two-qudit  state. As is known, the choice of measurement setting is important. There are measurement settings, for which even maximally entangled state may not violate an inequality. One of the key mathematical reason for the
relation (\ref{concurrence_entanglement}) to exist is that the sum (\ref{identity_3}) is independent
of $m$. In the case of CGLMP inequality, the function $f(\alpha)$ is different, therefore a different
sum occurs. That sum is not independent of $m$. Therefore, such a relation does not exist for
the  CGLMP inequality.  However the measurement of the Bell-SLK function can help us in
finding the amount of entanglement in a pure two-qudit state.

\section{Conclusion}
In conclusion, we have studied Bell-type inequalities for bipartite qudit systems. There are many
such local-realistic inequalities present in the literature. We have argued that one such inequality,
namely the SLK inequality, can be useful in measuring the entanglement present in such systems.
This also addresses an important question in entanglement theory: How to measure amount of 
entanglement for a bipartite state experimentally. The earlier methods for measuring entanglement requires number of observables to increase with dimension of the subsystems. In contrast, 
the scheme presented here requires only four observables to be measured to find the amount 
of entanglement present in a bipartite pure state. The scheme also works for a class of 
bipartite mixed qudit states -- isotropic states. We have also considered the case
of mixed states which are obtained after applying phase or amplitude damping channels 
on two qutrits which are in maximally entangled state. It appears that for such states one can
also measure entanglement, as characterized by negativity.
We have also discussed the experimental feasibility of our scheme.  
Current state-of-the-art allows 
experimental measurement of these observables (as in Eq. (\ref {settings})) with 
$d-1$ settings to measure $d$ possible outcomes
of each observable \cite{hplo}. Improvements in the experimental set up may allow the measurement of each
observable with fewer settings.

\section{Acknowledgment}

S.K.C. acknowledges support from the Council of Scientific and
Industrial Research, Government of India (Scientists' Pool Scheme).


\begin{thebibliography}{9}
\bibitem{standatom} If a priori, it is known that the state is pure, then we need only $2(d^2-1)$ measurements to reconstruct the state \cite{thew}.
\bibitem{footnote} The proof of the theorem can be extened for any nonintegral real value of $b$. The same result will hold.
      We note that the sum itself is not defined when $bk$ is an integer, as one of its term diverges. However, this
is not the case here, because here $bk=\frac{1}{4}$ which is not an integer.
\end{thebibliography}
\end{document}